\documentstyle[amsmath,epsf,a4,11pt]{article}

\begin{document}

\date{}
\title{{\bf G\"{o}del brane}}
\author{John D. Barrow$^{1}$\thanks{e-mail address:
j.d.barrow@damtp.cam.ac.uk} and Christos G. Tsagas$^{1,2}$
\thanks{e-mail address: c.tsagas@damtp.cam.ac.uk}\\
{\small $^1$DAMTP, Centre for Mathematical Sciences, University of
Cambridge,}\\ {\small Cambridge~CB3~0WA, UK}\\ {\small
$^2$Department of Mathematics and Applied Mathematics, University
of Cape Town,}\\ {\small Rondebosch 7701, South Africa}}

\maketitle

\begin{abstract}
We consider the brane-world generalisation of the G\"{o}del
universe and analyse its dynamical interaction with the bulk. The
exact homogeneity of the standard G\"{o}del spacetime no longer
holds, unless the bulk is also static. We show how the anisotropy
of the G\"{o}del-type brane is dictated by that of the bulk and
find that the converse is also true. This determines the precise
evolution of the nonlocal anisotropic stresses, without any
phenomenological assumptions, and leads to a self-consistent
closed set of equations for the evolution of the G\"{o}del brane.
We also examine the causality of the G\"{o}del brane and show that
the presence of the bulk cannot prevent the appearance of closed
timelike curves.\\\\ PACS number(s): 04.50.+h, 98.80.Cq
\end{abstract}


\section{Introduction}
The G\"{o}del universe~\cite{G,HE} has been a well-studied example
of the importance of understanding the global structure of
space-time. It sheds light on the impact of cosmological rotation,
the constraints imposed by space-time homogeneity, and the
unexpected compatibility between general relativity and the
presence of closed timelike curves. Whilst the G\"{o}del universe
is not a realistic model of our visible universe it may possess
ingredients which illuminate the structure of the actual universe,
and its stability properties reveal the extent to which its
unusual causal structure may be expected to appear as a component
of more general solutions of Einstein's equations. In this paper
we will investigate the structure of a G\"{o}del universe in a
braneworld model.

Developments in string theory have inspired the construction of
braneworld models, in which standard-model fields are confined to
a surface that constitutes our 3-brane universe, while gravity
propagates in all the spatial dimensions. A simple 5-D class of
such models allows for a non-compact extra dimension via a novel
mechanism for localization of gravity around the brane at low
energies. This mechanism is the warping of the metric by a
negative 5-D cosmological constant~\cite{RS}. These models have
been generalized to admit cosmological branes~\cite{SMS}, and they
provide an interesting arena in which to impose cosmological tests
on extra-dimensional generalizations of Einstein's
theory~\cite{M,MWBH}.

There are some interesting features of this scenario. An
anisotropic stress is imprinted on the brane by the 5-D Weyl
tensor, and this non-local field cannot be predicted by
brane-bound observers since it includes 5-D gravitational-wave
modes. The bulk equations are needed to determine the brane
dynamics completely~\cite{SMS}. The full 5-D problem also involves
choosing boundary conditions in the bulk. On the other hand,
starting from a brane-bound viewpoint, any choice of anisotropic
stress that is consistent with the brane equations will correspond
to a bulk geometry, which can be locally determined in principle
by numerical integration (or approximately, close to the brane, by
Taylor expanding the Lie-derivative bulk equations given
in~\cite{SMS}). However, numerical integration is very complicated
(see~\cite{CRSS} for the black hole case). Even if it can be
successfully performed, it will not give the global properties of
the bulk. The bulk geometry that arises for a given anisotropic
stress may have unphysical boundary conditions or singularities
(e.g., the bulk corresponding to a Schwarzschild black hole has a
string-like singularity and a singular Cauchy horizon~\cite{CHR}).
We have few exact bulk solutions to guide us in a study of
cosmological anisotropy. One known solution~\cite{BDEL} is the
Schwarzschild-AdS bulk that contains a (moving) Friedmann brane,
which is the exactly isotropic and homogeneous limit of an
anisotropic universe. Some examples of spatially homogeneous
Bianchi-type braneworlds have recently been given by Campos et
al~\cite{CMMS}.

In section 2 we derive the kinematical properties of a
non-expanding, shear-free, homogeneous, rigidly rotating brane of
G\"{o}del type. From these results we proceed in section 3 to
determine their consequences for the structure of the bulk. Unlike
the general situation, it is found that the G\"{o}del brane
creates a closed system of equations for the non-local bulk Weyl
stresses. Finally, in section 4 we discuss the causal structure of
the G\"{o}del brane and how closed timelike curves persist despite
the bulk effects.

\section{G\"{o}del brane}
We study the dynamics of the G\"{o}del universe in the context of
the brane-world scenario, by assuming the presence of a rigidly
rotating, non-expanding, shear-free and acceleration-free
space-time, containing a single perfect fluid on the brane.
Following~\cite{E}, these are the minimum conditions that
covariantly characterise the standard general relativistic
G\"{o}del model; we will also employ them to define what one might
call a G\"{o}del brane. In technical terms this means that, in
addition to the perfect fluid assumption, we will impose the
kinematical constraints $\Theta=0=A_{a}=\sigma_{ab}$,
$\omega_{a}\neq0$ and $\nabla_{a}\omega_{b}=0$ on the brane. The
scalar $\Theta$ describes the average volume expansion, $A_{a}$ is
the 4-acceleration, $\sigma$ is the shear rate and $\omega_{a}$ is
the vorticity rate associated with a chosen local timelike
4-velocity field $u_{a}$. Also, $\nabla_{a}$ is the covariant
derivative operator with respect to the brane metric $g_{ab}$. For
more details on the adopted formalism and notation the reader is
referred to~\cite{M}. Having defined the G\"{o}del brane, we can
use the fully nonlinear covariant relations of~\cite{M} to obtain
the equations that dictate the behaviour of the G\"{o}del universe
in the context of brane-world scenarios. Note that we do not
consider explicit expressions for the brane and the bulk metrics.
The embedding of braneworlds more complex than the original
Minkowski and the standard FRW branes has been the subject of
active research, but only a few complete anisotropic solutions are
available in the literature (see~\cite{CMMS} and references
therein). Constructing the metrics of both the Godel brane and of
the host 5-dimensional bulk is a question that goes beyond the
scope of this paper.

\subsection{Kinematics}
For a perfect fluid, the vorticity propagation and constraint
equations are trivially satisfied. Of the remaining non-trivial
kinematical relations, the generalised Raychaudhuri equation reads
\begin{equation}
{\textstyle{\frac{1}{2}}}\kappa^{2}(\rho+3p)- 2\omega^{2}-
\Lambda=-{\textstyle{\frac{1}{2}}}\kappa^{2}(2\rho+3p)
\frac{\rho}{\lambda}- \frac{6}{\kappa^{2}\lambda}{\cal U}\,,
\label{bRay}
\end{equation}
where $\kappa^{2}=8\pi/M_{P}^{2}$, ${\cal U}$ is the dark energy
density, $\lambda$ is the brane tension and the terms on the
right-hand side provide the local quadratic and bulk corrections.
The generalised shear propagation equation takes the form
\begin{equation}
E_{ab}+ \omega_{\langle a}\omega_{b\rangle}=
\frac{3}{\kappa^{2}\lambda}{\cal P}_{ab}\,,  \label{bsigmadot}
\end{equation}
where $E_{ab}$ is the magnetic component of the local Weyl tensor
and angled brackets indicate the symmetric and trace free-part of
tensors projected orthogonally to $u_{a}$ on the brane. Here the
corrections are due to the effective non-local anisotropic stress
${\cal P}_{ab}$ arising from the free gravitational field in the
bulk. Similarly, the generalised shear constraint reduces to
\begin{equation}
\frac{6}{\kappa^{2}\lambda}{\cal Q}_{a}=0 \Rightarrow {\cal Q}_{a}
=0\,.  \label{bsigmacon}
\end{equation}
This means that the effective non-local energy flux vanishes
identically, which reduces the available degrees of freedom in the
bulk. Finally, the generalised gravito-magnetic constraint and the
kinematics of the G\"{o}del space-time guarantee that the magnetic
part of the Weyl tensor vanishes:
\begin{equation}
H_{ab}=0\,,  \label{bHcon}
\end{equation}
exactly as in the standard G\"{o}del universe~\cite{E}.

\subsection{Conservation laws}
When the matter has a perfect fluid equation of state, with energy
density $\rho$ and pressure $p$, the local energy and momentum
conservation laws reduce to the constraints
\begin{equation}
\dot{\rho}=0\,,  \label{bedcon}
\end{equation}
and
\begin{equation}
{\rm D}_{a}p=0\,,  \label{bmdcon}
\end{equation}
respectively. Note that ${\rm D}_a=h_{a}{}^b\nabla_{b}$, with
$h_{ab}=g_{ab}+u_{a}u_{b}$, is the covariant derivative operator
orthogonal to $u_{a}$ on the brane. If the fluid is also
barotropic, with $p=p(\rho)$, the former of the above implies that
$\dot{p}=0$ and the latter that ${\rm D}_{a}\rho=0$. Therefore, as
in the standard G\"{o}del model, both $\rho$ and $p$ are
covariantly constant.

For the same matter content, and given the result
(\ref{bsigmacon}), the non-local conservation equations become
\begin{equation}
\dot{{\cal U}}=0\,,  \label{bbedcon}
\end{equation}
and
\begin{equation}
{\textstyle{\frac{1}{3}}}{\rm D}_{a}{\cal U}+ {\rm D}^{b}{\cal
P}_{ab}=0\,.  \label{bbmdcon}
\end{equation}
Therefore, the bulk dark energy density remains constant in time,
while its gradients are given by the transverse part of the
effective non-local anisotropic stresses.

\subsection{Ricci and Weyl curvature}
The orthogonally-projected Ricci tensor ${\cal R}_{ab}$ and Ricci
scalar ${\cal R}$, which are obtained from the generalised
Gauss-Codacci equation, satisfy the relations
\begin{equation}
{\cal R}_{\langle ab\rangle}=\frac{6}{\kappa^{2}\lambda}{\cal
P}_{ab}\,,  \label{bcR<ab>}
\end{equation}
and
\begin{equation}
{\cal R}= 2\left(\kappa^{2}\rho-\omega^{2}+\Lambda\right)+
\frac{\kappa^{2}\rho^{2}}{\lambda}+
\frac{12}{\kappa^{2}\lambda}{\cal U}\,.  \label{bcR}
\end{equation}
respectively. The latter is the generalised Friedmann equation on
the G\"{o}del brane.

The generalised propagation and constraint equations for the
electric and magnetic Weyl components, assuming that the local
matter is a barotropic fluid, are
\begin{equation}
\dot{E}_{ab}+ \omega^{c}\epsilon_{cd\langle a}E_{b\rangle }{}^{d}=
-\frac{3}{\kappa^{2}\lambda}\left(\dot{{\cal P}}_{ab}
+\omega^{c}\epsilon_{cd\langle a}{\cal
P}_{b\rangle}{}^{d}\right)\,, \label{bEdot}
\end{equation}

\begin{equation}
{\rm curl}E_{ab}=\frac{3}{\kappa^2\lambda}{\rm curl}{\cal
P}_{ab}\,,  \label{bHdot}
\end{equation}

\begin{equation}
{\rm D}^bE_{ab}=\frac{1}{\kappa^2\lambda}\left(2{\rm D}_a{\cal U}
-3{\rm D}^b {\cal P}_{ab}\right)=-\frac{9}{\kappa^2\lambda}{\rm
D}^b{\cal P}_{ab}\,,  \label{bdivE}
\end{equation}
with the latter equality obtained by means of (\ref{bbmdcon}), and
\begin{equation}
\kappa^2(\rho+p)\omega_a+ 3E_{ab}\omega^b=
-\frac{\kappa^2\rho}{\lambda}(\rho+p)\omega_a-
\frac{1}{\kappa^2\lambda}\left(8{\cal U}\omega_a-3{\cal
P}_{ab}\omega^b\right)\,.  \label{bdivH}
\end{equation}
Note that ${\rm curl}W_{ab}=\varepsilon_{cd\langle a}{\rm
D}^{c}W^{d}{}_{b\rangle}$ for every orthogonally projected
symmetric tensor $W_{ab}$ and $\varepsilon_{abc}$ is the projected
alternating tensor.

\subsection{Further constraints}
Given that $\omega _{a}$ is covariantly constant on the brane, the
projected divergence of Eq.~(\ref{bsigmadot}) means that
\begin{equation}
{\rm D}^{b}E_{ab}= \frac{3}{\kappa^{2}\lambda}{\rm D}^{b}{\cal
P}_{ab}\,.  \label{bdivE1}
\end{equation}
The above, with result (\ref{bdivE}), then imply that ${\rm
D}^{b}{\cal P}_{ab}=0$, which in turn leads to
\begin{equation}
{\rm D}_{a}{\cal U}=0\,,  \label{DacU}
\end{equation}
using the non-local momentum conservation law (\ref{bbmdcon}).
Combined with the bulk energy conservation law given by
(\ref{bbedcon}), this result guarantees that the non-local energy
of the G\"{o}del brane is covariantly constant and acts as a
(positive or negative) `dark' cosmological constant in, say,
Eq.~(\ref{bRay}).

Contracting (\ref{bsigmadot}) with $\omega_{a}$, substituting the
result into (\ref{bdivH}) and then contracting again with
$\omega_{a}$ we obtain
\begin{equation}
\kappa^{2}(\rho+p)- 2\omega^{2}=
-\frac{\kappa^{2}\rho}{\lambda}(\rho+p)-
\frac{8}{\kappa^{2}\lambda}{\cal U}-
\frac{6}{\kappa^{2}\lambda}{\cal P}_{ab}n^{a}n^{b}\,,
\label{baux1}
\end{equation}
where $n_{a}$ is the unit vector along the direction of the local
rotation axis (i.e.~$\omega_{a}=\omega n_{a}$, with $n_{a}u^{a}=0$
and $n_{a}n^{a}=1$). Substituted into (\ref{bRay}), this gives
\begin{equation}
\kappa^{2}(\rho-p)+ 2\Lambda= \frac{\kappa^{2}\rho p}{\lambda}-
\frac{4}{\kappa^{2}\lambda}{\cal U}-
\frac{12}{\kappa^{2}\lambda}{\cal P}_{ab}n^{a}n^{b}\,,
\label{baux2}
\end{equation}
which shows that, in contrast to the standard G\"{o}del
space-time, the local cosmological constant can now be positive.
For example, in the simple case where ${\cal U}=0={\cal P}_{ab}$
and $p=\rho$ we find that $\Lambda=\kappa\rho^{2}/2\lambda>0$.
Equations (\ref{baux1}) and (\ref{baux2}) combine to give
\begin{equation}
\kappa^{2}\rho- \omega^{2}+ \Lambda=
-\frac{\kappa^{2}\rho^{2}}{2\lambda}-
\frac{6}{\kappa^{2}\lambda}{\cal U}-
\frac{9}{\kappa^{2}\lambda}{\cal P}_{ab}n^{a}n^{b}\,,
\label{baux3}
\end{equation}
thus generalising the standard general relativistic constraints to
G\"{o}del branes (see~\cite{E}). Note that, on using the above,
the projected Ricci scalar (see Eq.~(\ref{bcR})) reduces to
\begin{equation}
{\cal R}=-\frac{18}{\kappa^{2}\lambda}{\cal P}_{ab}n^{a}n^{b}\,,
\label{bcR1}
\end{equation}
and ensures that ${\cal P}_{ab}$ determines both ${\cal
R}_{\langle ab\rangle}$ and ${\cal R}$.

Equations (\ref{bRay})-(\ref{bcR1}) determine the dynamics of the
G\"{o}del brane by incorporating the local and non-local higher
dimensional corrections. General relativity is recovered in the
low-energy limit $\lambda^{-1}\rightarrow0$. As it stands, the
system is not closed since there is no propagation equation for
the bulk anisotropic stress tensor ${\cal P}_{ab}$. The latter is
customarily determined phenomenologically, by imposing certain
physically and geometrically plausible constraints or by creating
an arbitrary propagation equation for ${\cal P}_{ab}$.

\section{Interaction between the brane and the bulk}
\subsection{Implications of the bulk}
The presence of the bulk in brane-world models means that some of
the standard G\"{o}del constraints do not hold any more. The
electric Weyl tensor, in particular, no longer satisfies the
simple constraints of the standard G\"{o}del space-time~\cite{E}.
This is clearly demonstrated by the non-local energy and stress
terms in Eqs.~(\ref{bsigmadot}) and (\ref{bEdot})-(\ref{bdivE}).
The latter equations ensure the dependence of the local tidal
forces on bulk quantities, such as ${\cal U}$ and ${\cal P}_{ab}$.
In fact, the time independence of the vorticity and
Eq.~(\ref{bsigmadot}) imply the relation
\begin{equation}
\dot{E}_{ab}=\frac{3}{\kappa^2 \lambda }\dot{{\cal P}}_{ab}
\label{bsigmadot1}
\end{equation}
between the time derivatives of $E_{ab}$ and ${\cal P}_{ab}$. This
underlines an interesting effect of the bulk: the stationary
nature of the G\"{o}del universe is not guaranteed unless the
non-local stresses are time-independent as well. This is also
clear from the dependence of the Ricci curvature on ${\cal
P}_{ab}$ (see Eqs.~(\ref{bcR<ab>}) and (\ref{bcR1})). Thus, both
the local tidal forces and the local curvature will vary in time
if the bulk anisotropy does so. Moreover, the presence of the bulk
can also affect the homogeneity of the G\"{o}del model, as
relation (\ref{bHdot}) verifies. Therefore, rigidly rotating,
non-expanding branes with zero shear and acceleration admit many
more possibilities than the limiting case of the general
relativistic G\"{o}del universe.

\subsection{Implications on the bulk}
The fact that the bulk can influence the standard symmetries of
the G\"{o}del universe, such as its stationary nature and spatial
homogeneity, is not entirely unexpected since the bulk introduces
additional degrees of freedom to the model. Analogous
modifications have also been identified in the brane-world
generalisation of the Einstein static universe for
example~\cite{GM}. What is rather surprising is that the remaining
symmetry in the G\"{o}del brane is enough to dictate the exact
behaviour of the bulk quantities, and in particular of the
non-local anisotropic stresses, in a self-consistent way. In
particular, consider the relation
\begin{equation}
\omega^{c}\epsilon_{cd\langle a}E_{b\rangle}{}^{d}=
\frac{3}{\kappa^{2}\lambda}\omega^{c}\epsilon_{cd\langle a}{\cal
P}_{b\rangle}{}^{d}\,,  \label{bsigmadot2}
\end{equation}
obtained by substituting $E_{ab}$ in the right-hand side of the
above from expression (\ref{bsigmadot}). Putting
(\ref{bsigmadot2}) back into Eq.~(\ref{bEdot}) and using
(\ref{bsigmadot1}) we arrive at the following simple evolution law
for the non-local anisotropic stresses
\begin{equation}
\dot{{\cal P}}_{ab}=-\omega^{c}\epsilon_{cd\langle a}{\cal
P}_{b\rangle}{}^{d}\,,  \label{bcPdot}
\end{equation}
which depends on the rotation of the local G\"{o}del model.
Accordingly, ${\cal P}_{ab}$ will remain zero if it is zero
initially, and the same also holds for ${\cal U}$ (see
Eq.~(\ref{bbedcon})). In that case the behaviour of the G\"{o}del
brane approaches that of the standard G\"{o}del model. Note that
by substituting the above result into Eq.~(\ref{bEdot}), the
evolution law of the electric Weyl component reduces to
\begin{equation}
\dot{E}_{ab}=-\omega^c\epsilon_{cd\langle a}E_{b\rangle}{}^d\,,
\label{bEdot2}
\end{equation}
ensuring the consistency of the system
(\ref{bsigmadot1})-(\ref{bEdot2}).

The existence of the evolution law (\ref{bcPdot}) means that the
G\"{o}del brane is described by a closed and consistent set of
equations, which incorporates the effects of the bulk anisotropy,
without the need of extra phenomenological assumptions. This is a
consequence of the time independence of the local vorticity
vector, which leads to Eq.~(\ref{bsigmadot1}) first and then to
the propagation equation (\ref{bcPdot}). Obviously, the latter no
longer holds if we relax the symmetry by allowing
$\dot{\omega}_{a}\neq 0$.

\section{Causality in the G\"odel brane}
In the last section we showed how the presence of the bulk can
affect basic features of the G\"{o}del universe, such as its
stationarity. In view of this, it is worth examining whether the
non-local effects can affect another key property of the G\"{o}del
model, namely the presence of closed timelike curves. This can
happen in general relativistic G\"{o}del-type cosmologies, that is
in rigidly rotating homogeneous spacetimes with extra matter
sources due to higher-order terms in the Lagrangian and in the
presence of some string theory corrections~\cite{RT,A}.

The induced field equations on the brane comprise a modified set
of the standard Einstein equations with new terms that carry the
bulk effects onto the brane~\cite{SMS,M}
\begin{equation}
R_{ab}-{\textstyle{1\over2}}Rg_{ab}=\kappa^2T_{ab}+
\frac{6\kappa^2}{\lambda}S_{ab}- {\cal E}_{ab}- \Lambda g_{ab}\,,
\label{bEFE1}
\end{equation}
where the tensors $S_{ab}$ and ${\cal E}_{ab}$ respectively
describe the local and nonlocal higher dimensional corrections. In
our case $T_{ab}$ has the standard perfect fluid form, which means
that
\begin{equation}
S_{ab}={\textstyle{1\over12}}\rho^2u_au_b+
{\textstyle{1\over12}}\rho(\rho+2p)h_{ab}\,,  \label{Sab}
\end{equation}
with $S=S_a{}^a=\rho(\rho+3p)/6$. Also,
\begin{equation}
{\cal E}_{ab}=-\frac{6}{\kappa^2\lambda}\left[{\cal
U}\left(u_au_b+{\textstyle{1\over3}}h_{ab}\right)+{\cal
P}_{ab}\right]\,, \label{cEab}
\end{equation}
with ${\cal E}={\cal E}_a{}^a=0$. For a homogeneous brane the bulk
has no temporal or spatial dependence either (see
Eqs.~(\ref{bHdot}), (\ref{bdivE}) and (\ref{bsigmadot2})). We can
therefore treat this brane-bulk configuration as a homogeneous
G\"{o}del-type spacetime where the bulk acts as an extra source to
the energy-momentum tensor of the matter~\cite{RT}. The latter is
given by
\begin{equation}
{\cal T}_{ab}=T_{ab}+ \frac{6}{\lambda}S_{ab}-
\frac{1}{\kappa^2}{\cal E}_{ab}\,,  \label{cTab}
\end{equation}
where ${\cal T}={\cal
T}_a{}^a=3p(1+\rho/\lambda)-\rho(1-\rho/\lambda)$, while the
associated field equations read
\begin{equation}
R_{ab}=\kappa^2{\cal T}_{ab}- {\textstyle{1\over2}}\kappa^2{\cal
T}g_{ab}+ \Lambda g_{ab}\,, \label{bEFE2}
\end{equation}
given that $R=4\Lambda-\kappa^2{\cal T}$.

Adopting the signature conventions of~\cite{RT} and using
cylindrical coordinates, the metric of the aforementioned
G\"{o}del-type spacetime has the general form
\begin{equation}
{\rm d}s^{2}=\left[ {\rm d}t+\frac{4\omega }{m^{2}}\sinh
^{2}\left( \frac{m\,r}{2}\right) {\rm d}\phi \right]
^{2}-\frac{\sinh ^{2}(m\,r)}{m^{2}}{\rm d}\phi ^{2}-{\rm
d}r^{2}-{\rm d}z^{2}\,, \label{metric}
\end{equation}%
where the parameter $m$ specifies the class of the solution
associated with the G\"{o}del-type spacetime and also determines
the causality of the model~\cite{RT}. In particular, there will be
no closed timelike curves if $m^2\geq4\omega^2$. Otherwise
causality is not preserved. Following~\cite{RT}, we introduce an
orthonormal frame relative to which the field equations
(\ref{bEFE2}) read
\begin{equation}
R_{AB}={\cal T}_{AB}- {\textstyle{\frac{1}{2}}}\kappa^2{\cal
T}\,n_{AB}- \Lambda\,n_{AB}\,, \label{RAB}
\end{equation}
where $n_{AB}$ is the Minkowski metric with signature
-2.\footnote{The reader is referred to~\cite{RT} for further
details on the formalism and the notation used in this section.}
In the same frame the energy-momentum tensor of the effective
fluid which incorporates the bulk effects is
\begin{equation}
{\cal T}_{AB}=\left[\rho\left(1+\frac{\rho}{2\lambda}\right)+
\frac{6{\cal U}}{\kappa^4\lambda}\right]V_A\,V_B-
\left[p+\frac{\rho(\rho+2p)}{2\lambda}+\frac{2{\cal
U}}{\kappa^4\lambda}\right]h_{AB}+ \frac{6}{\kappa^4\lambda}{\cal
P}_{AB}\,, \label{cTAB}
\end{equation}
with $V^A=\delta_0{}^A$ and $h_{AB}=n_{AB}-V_AV_B$. Substituting
the above into Eq.~(\ref{RAB}) and then employing a
straightforward calculation we obtain
\begin{equation}
R_{00}={\textstyle{1\over2}}\kappa^2\rho\left(1+\frac{2\rho}{\lambda}\right)+
\frac{6{\cal U}}{\kappa^2\lambda}+
{\textstyle{3\over2}}\kappa^2p\left(1+\frac{\rho}{\lambda}\right)-
\Lambda\,,  \label{R00}
\end{equation}
and
\begin{equation}
R_{11}={\textstyle{\frac{1}{2}}}\kappa^2(\rho-p)-
\frac{\kappa^2\rho p}{2\lambda}+ \frac{2{\cal
U}}{\kappa^2\lambda}+ \frac{6}{\kappa^2\lambda}{\cal P}_{11}+
\Lambda\,, \label{R11}
\end{equation}
with exactly analogous expressions for $R_{22}$ and $R_{33}$. In
addition, $R_{02}=0$ which ensures that $\omega$ is
constant~\cite{RT}. The latter agrees with our original assumption
of a rigidly rotating space. Finally, from~\cite{RT} we have
\begin{equation}
R_{00}=2\omega ^{2}\hspace{10mm}{\rm
and}\hspace{10mm}R_{11}=R_{00}-m^{2}\,, \label{R00-R11}
\end{equation}
where the parameter $m$ determines the class of the G\"{o}del-type
solution associated with our model. Results (\ref{R11}) and
(\ref{R00-R11}) combine to give
\begin{equation}
m^{2}=2\omega^2-{\textstyle{1\over2}}\kappa^2(\rho-p)+
\frac{\kappa^2\rho p}{2\lambda}- \frac{2{\cal
U}}{\kappa^2\lambda}- \frac{6}{\kappa^2\lambda}{\cal P}_{11}-
\Lambda\,.  \label{m2}
\end{equation}
At first it might seem that for certain brane-bulk configurations
the right-hand side should be able to reach the desired value of
$4\omega^2$ in which case the causality of the associated
G\"{o}del-type brane would have been preserved. On using the
constraint (\ref{baux2}), however, the above reduces to
\begin{equation}
m^2=2\omega^2+ \frac{6}{\kappa^2\lambda}\left({\cal
P}_{ab}n^an^b-{\cal P}_{11}\right)\,.  \label{m2*}
\end{equation}
Then, since we can always align our frame so that one of its
spatial axes is in the direction of the vorticity vector, we have
${\cal P}_{11}={\cal P}_{ab}n^an^b$. This implies that
$m^2=2\omega^2$ just like in the standard G\"{o}del model. In
other words, the presence of the bulk cannot affect the causality
of the brane and closed timelike curves still exist.

\section{Discussion}
G\"{o}del's rotating cosmos has been one of the most intriguing
solutions of the Einstein field equations. Since its discovery in
1949, the G\"{o}del model has attracted continual attention
because it contains closed timelike curves. This completely
unexpected discovery not only stimulated the rigorous
investigation of the global structure of spacetimes, but led to a
reappraisal of our thinking about issues such as Mach's Principle
and the possibility of time travel within relativistic theories.
It has also led to a series of G\"{o}del-type solutions
\cite{RT}-\cite{RaSh} produced by the introduction of extra fields
or of quantum-gravity corrections. Interestingly, in some of these
solutions the causal pathologies usually associated with the
G\"{o}del universe do not occur.

It is generally believed that general relativity breaks down at
very high energies and it is likely to be the limit of a more
general theory. Developments in string theory suggest that gravity
may be higher dimensional, confined to a 4-dimensional part of
space-time at low energies. A number of these theories confine the
matter to a 3-brane, while the gravitational field is free to
propagate in the extra dimensions. In a popular realisation of
this scenario, gravity is localised on a single 3-brane embedded
within a 5-dimensional \textquotedblleft bulk\textquotedblright\
space. This idea has triggered much work on the properties of
these braneworld models, on their differences from their general
relativistic counterparts, and on their potential for experimental
test. Here, we have investigated the nature of a braneworld that
exhibits the structure of the G\"{o}del universe. Interesting
differences emerge compared to the usual G\"{o}del universe of
general relativity due to the influence of the bulk stresses. The
exact homogeneity of the G\"{o}del space-time, for example, is no
longer preserved unless the bulk is also static. Also, in contrast
to the general braneworld case, the evolution formulae of the
G\"{o}del brane form a closed set of equations. In particular, the
G\"{o}del constraints enforce a single propagation equation for
the non-local anisotropic stresses, which generally are determined
on purely phenomenological grounds. We have also considered the
causality of the G\"{o}del brane by treating it as an effective
G\"{o}del-type spacetime where the bulk effects are represented as
extra source terms in the energy momentum tensor of the matter. In
particular, using the techniques of~\cite{RT} we were able to show
that the presence of the bulk does not affect the causality of the
model and that closed timelike curves still exist. Note that
analogous studies of G\"{o}del-type spacetimes with extra sources,
in the form of a scalar field or of higher-order gravity
corrections, showed that in certain cases the causality problems
of the general relativistic G\"{o}del cosmos can be avoided. Our
analysis, however, shows that this not the case when dealing with
the G\"{o}del-type brane. We attribute this negative result to the
highly constraint bulk of the G\"odel braneworld. It is the strict
symmetries of the G\"{o}del brane which specify the properties of
the bulk, leaving no residual freedom that can be used to remove
the causality pathologies of the model.

\section*{Acknowledgements}
CT was supported by a Sida/NRF grant, PPARC and DAA. He also
wishes to thank the Centre for Mathematical Sciences at DAMTP,
where this work started, for their hospitality.

\end{document}